# Can Twitter Give Insights into International Differences in Covid-19 Vaccination? Eight countries' English tweets to 21 March 2021

Mike Thelwall, Statistical Cybermetrics Research Group, University of Wolverhampton, UK. Orcid: 0000-0001-6065-205X

Vaccination programs may help the world to reduce or eliminate Covid-19. Information about them may help countries to design theirs more effectively, with important benefits for public health. This article investigates whether it is possible to get insights into national vaccination programmes from a quick international comparison of public comments on Twitter. For this, word association thematic analysis (WATA) was applied to English-language vaccine-related tweets from eight countries gathered between 5 December 2020 and 21 March 2021. The method was able to quickly identify multiple international differences. Whilst some were irrelevant, potentially non-trivial differences include differing extents to which non-government scientific experts are important to national vaccination discussions. For example, Ireland seemed to be the only country in which university presidents were widely tweeted about in vaccine discussions. India's vaccine kindness term #VaccineMaitri was another interesting difference, highlighting the need for international sharing.

## Introduction

Vaccination on an international scale seems essential to eliminate Covid-19 or reduce its lethality. Although the World Health Organisation (WHO) and some international bodies like the EU are helping to organise vaccination drives, the logistics seem likely to be primarily organised by countries and reactions to them may have national cultural dimensions. In this context, international comparisons of vaccination strategies and public reactions to them may be useful to help identify common concerns, unusual issues and other factors that disproportionately affect individual nations. This is an urgent problem given the high Covid-19 international daily death rates.

Many international differences in Covid-19 vaccination needs are obvious and do not need a Twitter analysis. For example, countries differ in the number of vaccines and the health infrastructure available for vaccination programmes. Previous public health research about Covid-19 vaccination programmes has tended to focus on the threat posed by vaccine hesitancy (Paul et al., 2021; Malik et al., 2020) rather than wider public perceptions. National vaccination campaigns have been summarised in letters to editors (Bagcchi 2021; Sah et al., 2021) and the planning process has also been discussed, but focusing on vaccine hesitancy (DeRoo et al., 2020). Early vaccination demographics have also been reported (Painter et al., 2021). Vaccination passports are another important issue, including their legal implications (Phelan, 2020). The order of priority for vaccinations may be particularly relevant to members of the public and can influence death rates. There have been proposals about how this should be decided for individuals (Bell et al., 2020; Hassan-Smith et al., 2020) and bulk supplies (McClung, et al., 2020). No academic publication seems to have reported public attitudes towards Covid-19 vaccination from a general perspective however, whether on social media or through questionnaires, and few public health vaccination papers have focused on international comparisons.

The current study examines one aspect of the human dimension of public reactions towards Covid-19 vaccines: their discussion on Twitter. Twitter was chosen because it is a public site, in contrast to much of Facebook and Instagram, and allows researchers to gather tweets in real time. It appears to be the largest multinational free source of English-language public comments in the social web and was chosen for this reason. Any analysis of Twitter is biased towards the demographic that uses it, but this an unavoidable trade-off for a fast, low-cost analysis. It seems likely that older, less educated, and less technologically aware citizens tweet less (Blank et al., 2020; Wojcik & Hughes, 2019). The tweets from each country were analysed with Word Association Thematic Analysis (WATA), which is designed to identify themes in the differences between two sets of texts (Thelwall, 2021). The method draws attention to these themes that the analyst must then explain, potentially drawing attention to differences that could otherwise be overlooked (Thelwall et al., in press). WATA is more directly relevant than other exploratory text analysis methods that do not focus on difference, such as topic modelling (Ramage et al., 2009). It is chosen in preference to corpus linguistics methods that compare subsets of texts (Kennedy, 2014) because of its use of statistical thresholds and the availability in the integrated software Mozdeh that can also collect tweets. The aim of this paper is to test whether WATA could identify a set of vaccine-related topics that differ internationally on English-language Twitter.

## Methods

The research design was to (a) obtain a large sample of English-language tweets related to Covid-19 vaccination campaigns, (b) apply the WATA method to detect national differences, then (c) make qualitative judgements about whether the differences found are potentially useful from a public health perspective.

### Data: Vaccine-related tweets from multiple countries

Vaccine-related tweets were gathered from the Twitter Applications Programming Interface (API) in the free software Mozdeh with the following single keyword queries: vaccine, vaccination, vaccinating, vaccinated, CovidVaccination, CovidVaccine, CovidVaccineFacts. These queries were constructed based on test searches for vaccine-related keywords on the 5[th] of December 2020 and may favour terms that were more popular on that day. Anti-vaccination terms were not included because vaccination rather than vaccine hesitancy was the topic investigated. No terms specifying Covid-19 were added to the first four queries since Covid-19 vaccine seemed to be the only type discussed on Twitter on the day that the collection started, so the extra terms seemed unnecessary and would reduce the number of tweets collected. The queries were submitted at the maximum rate permitted by the Twitter API from 5 December 2020 to 21 March 2021. This arbitrary period was sufficient to obtain substantial numbers of tweets from multiple countries. English was specified as the language for the matching tweets.

Twitter can contain multiple copies of the same tweet due to retweeting and copy tweeting, as well as many very similar tweets from bots. To reduce the influence of these, the tweets were filtered to remove duplicates and near duplicates (tweets with identical content to another tweet after removing hashtags and @usernames).

The method used here to detect themes can be influenced by prolific tweeters so to reduce the influence of individuals, tweeters were allowed a maximum of one tweet per month. Anyone with multiple tweets in a single calendar month had all except one (chosen with a random number generator) deleted. Thus, no tweeter has more than four tweets in the full dataset. This left 5,789,082 tweets from the original set of 60,481,596.

Tweeters were assigned to countries based on the self-reported location information in their Twitter biography, if any. Many users reported a city or town instead of a country and these were matched against large city names that map to single countries to make additional country assignments. For example, a user from "Lagos" would be assigned to Nigeria. This information may be inaccurate if, for example, a user reports their birth country or ethnic origin rather than current location but it seems likely to be broadly accurate. As described below, the eight countries with the most tweets were selected for analysis.

## Analysis

WATA was applied to the tweets collected. For each country, the software Mozdeh was used to detect words that are statistically significantly more common in one of the eight countries investigated than in the remaining tweets (Figure 1). This automatic stage is then followed up by a manual stage, Word Association Contextualisation (WAC), which involves reading a sample of tweets from the selected country containing the term to identify its context. The top 100 terms from each country were analysed for this (800 in total), all of which were statistically significant (using a 2x2 chisquare test with familywise error rate correction: Benjamini & Hochberg, 1995). Only the top 100 terms were investigated to focus on the most substantial differences and as a practical consideration for a quick method. The WAC stage was applied in parallel to the eight selected countries.

Thematic Analysis (TA) was applied to the contextualised words to fit them into themes relevant to the research goals. Each theme is a way to generalise the contexts of multiple words. For example, the Geographic names theme stemmed from the observation that many terms were country or place names, so these could reasonably be grouped together into a single theme. The TA part of WATA is like traditional reflexive TA (Braun & Clarke, 2013) except that words are analysed rather than texts and each word is given a single theme. Words matching multiple contexts or themes were ignored. This stage was conducted in parallel for the eight countries, using the same themes for all countries so that the results could be compared. Two rounds of coding were used, first to generate the themes and second to check that each word was in an appropriate theme. Whilst TA usually requires additional round for cross-checking classifications and rearranging themes, to speed the process this was only done once. Whilst all the themes in a traditional TA are equally relevant, in this case themes that are not relevant to vaccination programmes, such as Geographic names, are irrelevant and primarily serve to filter out uninformative word differences.

Terms that did not fit easily into themes were recorded to be discussed separately because the purpose of the study is to identify international differences. Follow-up analyses were also conducted to investigate differences identified by the WATA.

The WATA was conducted by a single coder for speed because of the time-sensitive nature of the topic. This reduces the robustness of the results compared to the more standard approach of using multiple coders because the themes are more likely to be subjective. These themes can be checked in the online supplement, however (https://doi.org/10.6084/m9.figshare.14308490).

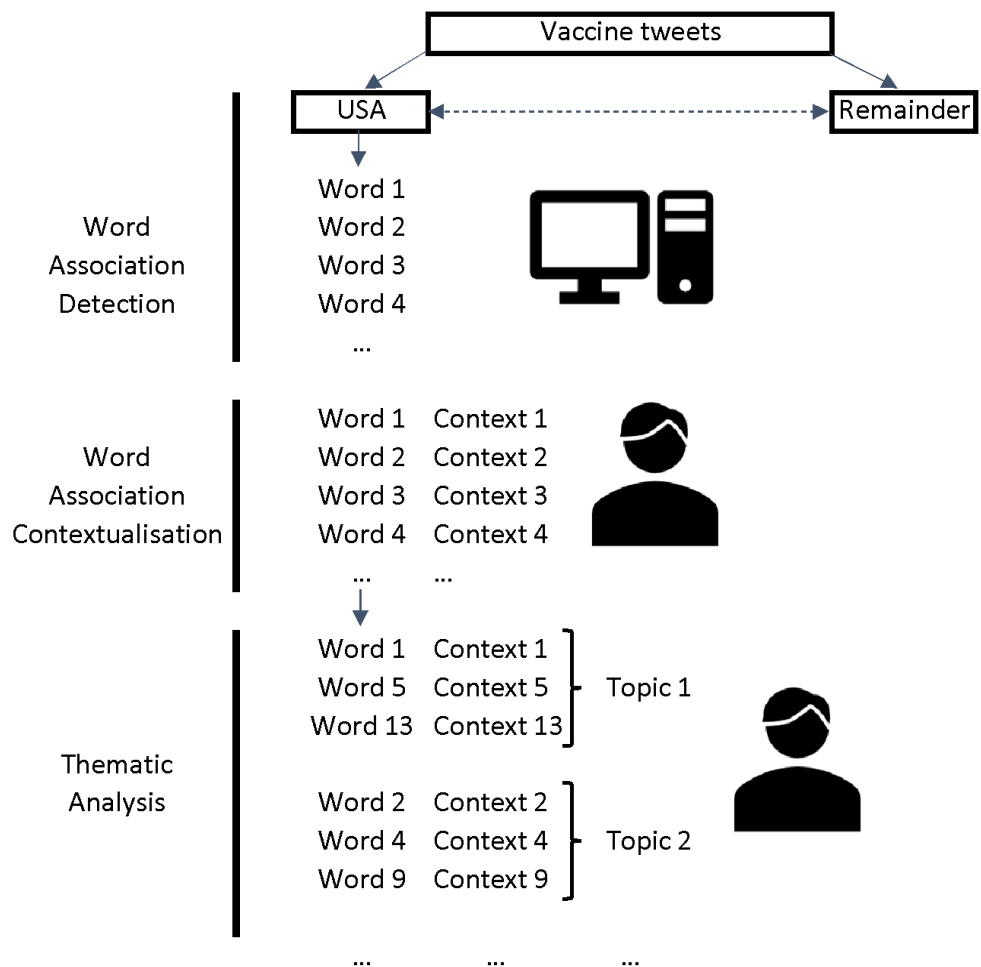

Figure 1. The WATA stages for the USA. The WATA process was conducted in parallel for the other seven countries.

## Results

The data collection produced 5,789,082 non-duplicate tweets matching the vaccination queries between 5 December 2020 and 21 March 2021, at a maximum of one tweet per user per month. Just under half of the tweeters in the vaccine tweets collection declared a country location or major city, with 223 countries represented. English is a major language of the top eight, but not the ninth country (Germany) so the top eight were selected for analysis (Table 1). This cut-off is relatively arbitrary, but whilst English is the official language of Nigeria and Nigerian English is widely used to communicate between people from different language groups (e.g., Hausa, Yoruba), it is not an official language of Germany, and is a joint official language in Pakistan (with Urdu). Tweeting in English may associate with being more educated in both Germany and Pakistan, so it seems reasonable to exclude both. The eight countries cover five continents and in the United Nations 2020 Human Development Index (http://hdr.undp.org/en/2020-report) are ranked very high, high (South Africa), Medium (India) or Low (Nigeria).

Table 1. The top ten countries declared by users tweeting about vaccines in English.

| Tweeters | % | Tweeter country |
|---|---|---|
| 3080305 | 53.2% | None declared |
| 1529220 | 26.4% | USA |
| 505756 | 8.7% | UK |
| 158537 | 2.7% | Canada |
| 114870 | 2.0% | India |
| 50535 | 0.9% | Australia |
| 45618 | 0.8% | South Africa |
| 34036 | 0.6% | Ireland |
| 28690 | 0.5% | Nigeria |
| 16660 | 0.3% | Germany |
| 14154 | 0.2% | Pakistan |

A quick gender analysis of the tweets is reported in the appendix, for information.

## International difference vaccine themes on Twitter

The following themes were identified in the top 100 words disproportionately occurring in the English vaccine tweets of each country compared to all other countries. The words and their organisation into themes can be checked in the online supplement (https://doi.org/10.6084/m9.figshare.14308490). Most terms were of types that are of little relevance to the current international drive to vaccinate the world, but the last three types have some relevance and are discussed in more detail. These may shed light on how the people in each country judge the arrangements to deliver vaccination programmes and the non-media sources of vaccine-related information consulted.

- **Geographic names**: Country and city names were mentioned enough in the tweets to appear widely for all countries. Such names are normally distinctive, with a few exceptions such as the London in the UK and Canada.
- **Local language or slang**: National slang words or word variants were most prevalent in Nigeria because of Nigerian English but also occurred due to national word preferences or slang (e.g., UK citizens have a vaccine jab, US citizens get a vaccine shot, and Nigerians collect the vaccine).
- **Politics, politicians or government** (other than health departments): Politicians and (to a lesser extent) political party names differ between countries and so those making contributions about vaccines are a natural inclusion in the lists.
- **News sources**: News organisation and journalists tend to be primarily national or at least have national brand names. This category also includes social media influencers that tweet extensively about news-related topics. There is a fuzzy boundary between social media active politicians and influencers, but influencers were classed as politicians when they declared on official political post.
- **Health services**: The organizations delivering health care differ between countries.
- **Vaccine names or manufacturers**: The vaccines developed or deployed varied between countries.
- **Lockdown**: Tweets from the UK and USA sometimes mentioned the need for social distancing and masks or discussed when social distancing restrictions might ease due to vaccination.

- **Vaccine rollout arrangements** (except for selecting participants): Countries differ in the details of where and how vaccines are administered. These seemed to be employer-based and online booking through a pharmacy chain (CVS) in the USA, whereas people in the UK tweeted about being invited by letter. Tweeters in India mentioned their Co-WIN online portal for booking, and from Nigeria the National Identification Number was mentioned as a possible route to access vaccines. Some Irish tweets mentioned the use of the Defence Forces to help the vaccination initiatives.

- **Qualifying for a vaccine**: The order in which people get vaccinated may be affected by different factors between nations, such as price, health condition, and job. In the US, the groups mentioned for priority include health workers, essential workers, educators, and older folk. UK tweets mentioned clinically vulnerable groups, carers, health staff, and older age groups. In Canada indigenous peoples were mentioned as needing special vaccination steps, and those in Long Term Care (LTC) were listed as a priority. Irish tweets mentioned the need to vaccinate family carers. No special groups were mentioned for India, Australia, or South Africa.

- **Medical experts, virologists or other COVID-19-related experts**: People being tweeted about in a non-official capacity but within the realm of their COVID-19-related expertise were classed in this group. For example, an epidemiologist tweeting her epidemiological contributions would fit in this group, but a mathematician tweeting her opinions about virology would be classed as a news source (influencer) instead. A prominent unofficial expert may reflect distrust of national strategies.
  - In the USA there were no prominent experts in the top 100 terms or even the top 500 terms. Media terms (e.g., CNN, #FoxNews) were also absent, although this may be partly due to US media being quoted around the world, lowering its word frequency test score.
  - In the UK, there were no experts in the top 100 terms. The top independent expert was Professor & Chair of Global Public Health, Edinburgh University Medical School Prof. Devi Sridhar (@devisridhar) at term position 174. She has provided consistent commentary on Covid-19, has argued for an eradication strategy in the UK, has been interviewed by the media and has given evidence to UK government groups.
  - Canada had the most COVID-19 experts (five) in the top 100 terms. One was Dr. Jennifer Kwan (@jkwan_md), giving a family physician's perspective. She also gives media interviews and has a focus on Ontario.
  - There were no health experts in the top 100 terms for India, with the highest ranked (310) being Faheem Younus, MD, Chief of Infectious Diseases at the University of Maryland in the USA. He received many questions on Twitter, with three times as many from India as any other country, with Pakistan being second. He tweets have an international flavour.
  - Australia had three COVID-19 experts in the top 100 terms. Broadcaster and General Practitioner Dr Vyom Sharma (@drvyom) was the first, tweeting about health issues related to Covid-19.
  - South Africa had one Covid-19 expert in the top 100. Prof Mosa Moshabela (@MoshabelaMosa), Deputy Vice-Chancellor at the University of KwaZulu-Natal, is a medical specialist and trained medical practitioner. He tweets public health messages related to Covid-19 and has been interviewed on television.

- o Ireland had four COVID-19 experts in the top 100 terms. The first was university immunologist Prof. Luke O'Neil, tweeting general Covid-19 science and information. Two university presidents, Philip Nolan of Maynooth University and Brian MacCraith of Dublin City University focused on the science of Covid-19, perhaps playing the role of filtering high quality scientific information to inform citizens and politicians in Ireland.
- o Nigeria had two COVID-19 experts in the top 100 terms, starting with Dr Chinonso Egemba (brand: The Aproko Doctor), who specialises in "leveraging social media for health education" (https://aprokodoctor.com/).

## *National vaccination tweeting topics*

The main distinctive features of the top 100 term list for each country are discussed separately below, mentioning the main themes above when relevant. Terms outside all themes are also introduced.

- **USA**: The US had 16 terms from people reporting that they or someone else had been vaccinated, forming the most distinctive trend. No other country had terms about the need for vaccination before in-person working, and only Australia also had a term related to the need for Covid-19 economic relief. Vaccine hesitancy in the Black community was also mentioned.
- **UK**: The UK had health service terms at the top of its list, reflecting a single public body in charge of health in the country. The most distinctive theme was the use of positive terms, however. These included fantastic, lovely, brilliant, pleased, well done, amazing, x and xx. The US had the term excited (to get a vaccine) and Ireland had delighted (to get a vaccine). Other UK issues included whether vaccine passports would be necessary for certain activities, when it would be possible to ease the lockdown (aided by vaccination), and whether the UK Test and Trace system was an expensive failure. Vaccine hesitancy in the Black And Minority Ethnic (BAME) community was also mentioned.
- **Canada**: Most of the top words were geographic names or codes, politicians or media sources. The relatively large number was due to the prominence of state names, news sources and politicians in addition to national politicians and news sources. The two vaccine qualification issues mentioned were the LTC (long-term care) priority and vaccine mistrust amongst indigenous Canadians.
- **India**: Many terms from India were geographic or language-related, partly due to state-based differences in COVID-19 strategies and politicians. India included the most terms of any country (10) related to vaccination rollout arrangements. This was partly caused by the public dry run of the vaccination programme, in advance of delivery, and partly by publicity for India running the world's largest vaccination drive (Bagcchi, 2021). India was the only country with a vaccine sharing term: #VaccineMaitri. This hashtag might be translated as vaccine kindness and was used in stories about India's powerful pharmaceutical industry shipping vaccines to other countries.
- **Australia**: A uniquely Australian topic was the question of whether vaccination would be required for arriving overseas travellers. There were also terms related to the Australian Open tennis tournament.
- **South Africa**: Possible corruption in Covid-19-related pay-outs was a unique topic for South Africa. Although corruption has also been alleged in the UK, no associated terms were in the top 100 UK words. Another issue discussed was that a batch of AstraZeneca vaccine might not be able to be used (less effective against the South African virus variant) or exchanged with another country because of its short expiry date. Two businesspeople were also frequently mentioned for their vaccination related comments: Both Khandani Msibi and Unathi Kwaza acted as social media influencers, tweeting opinions about the news.

- **Ireland**: The Irish list included two entrepreneurs, Pat Phelan and Rory McEvoy. Only South Africa also included businesspeople. All were acting as influencers, commenting in the media and on Twitter about the pandemic to reasonably large numbers followers.
- **Nigeria**: Two themes were in the top terms for Nigeria alone. First, shortages of palliative medicines for people with Covid-19 was commented on in some vaccine tweets, connecting the two (e.g., they hid the palliative medicines so will they hide the vaccines?) The Nigerian terms also included malaria from tweets calling for vaccination for this disease. This was the only issue raised that is unrelated to Covid-19.

## Discussion and conclusions

This study has multiple limitations. It is limited to a single language, social media service and period and the results are subjective. International differences fot other countries (e.g., China, Vietnam, Brazil, Russia) would not have been found, limiting the generalisability of the findings. Moreover, the method cannot easily be extended to compare between countries using different languages. The word frequency method used may overlook important issues that lack distinctive terms, so the absence of themes should not be taken as evidence of their irrelevance to a country. Another limitation is that the results have not been validated for usefulness, for example by consulting experts about whether they would be informative.

The main conclusion is that the method was able to quickly identify some meaningful differences between countries. Whilst many of the themes reflecting differences were irrelevant to Covid-19 vaccination programmes (e.g., slang, geographic names, media and politicians' names), and others provided a reminder of structural differences in programmes that an expert might already know (e.g., different vaccine priority strategies, different vaccination infrastructures, the vaccine passport issue, apparently failed test and trace systems, corruption allegations), a few suggested other underlying differences that may be of interest, as follows.

The differing relevance of independent medical experts is a potential concern given that coherent public health messages are necessary for vaccination programs. Whilst prominent independent experts suggest a vacuum for medical information, which would be a concern for official health channels in each country, there are other plausible explanations, such as celebrity scientist media cultures or local cultures of health information consumption. Similarly, the absence of prominent independent health experts does not indicate national consensus about vaccination, as in the case of the USA with a high level of vaccine hesitancy (Funk & Tyson, 2020) but no independent health expert in the tweet terms examined. The presence of non-scientific influencers from the business community in Ireland and South Africa is a potential cause for concern on this sensitive topic, although mainstream news sources may also be non-expert influential commentators. Finally, from India, the vaccine kindness concept, #VaccineMaitri, also seems important from a global perspective, especially if having a named concept encourages a more equitable international sharing of supplies.

Although it is outside the scope of this quick reaction paper to judge whether the differences found are insightful for policy makers, this paper has established a practical free, quick technique for getting international insights into public reactions to vaccination programmes. It seems possible that some of the differences summarised in the above paragraph might be of interest from a public health perspective within some countries.

## *Appendix: Gender differences*

The tweets were also analysed for evidence of differences in gender perspectives between countries by finding terms associated with either males or females, using the WAD word frequency approach above. The results are available in the online supplement (https://doi.org/10.6084/m9.figshare.14308490) but do not give new insights into Covid-19 reactions. In brief, tweeter genders were guessed using lists of commonly gendered first names in each country from gender-api.com (at least 90% used by males or females, ignoring people with other names). This produced enough gendered names for analysis in all countries with male or female gendered terms were found for the USA (1360 statistically significant terms), the UK (1125), Canada (221), Ireland (87), India (52), Australia (27), Nigeria (4) and South Africa (2). Nonbinary genders were detected from they/them pronouns listed in Twitter biographies but there were too few to analyse statistically.

Two gendered themes emerged. Except for Nigeria, females focused more on family members (reporting their vaccination). Except for Australia, Nigeria and South Africa, males focused more on politics. The exceptions may have been due to less data in these cases rather than a lack of underlying gender differences.